\documentclass[preprint,aps,superscriptaddress,amsmath,amssymb,nofootinbib]{revtex4-1}
\usepackage{dcolumn}
\usepackage{graphicx}
\usepackage{mathrsfs}
\usepackage{float}
\usepackage{physics}
\usepackage{xcolor}
\usepackage[colorlinks=true,allcolors=blue]{hyperref}

\begin{document}
\title{Linear Structure, Convergence and Resurgence of Hydrodynamic Transseries}

\author{Qi Zhou}
\email{qizhou@m.scnu.edu.cn}
\address{State Key Laboratory of Nuclear Physics and Technology, Institute of Quantum Matter, South China Normal University, Guangzhou 510006, China}
\address{Guangdong Basic Research Center of Excellence for Structure and Fundamental Interactions of Matter, Guangdong Provincial Key Laboratory of Nuclear Science, Guangzhou 510006, China}

\author{Enke Wang}
\email{wangek@scnu.edu.cn}
\address{State Key Laboratory of Nuclear Physics and Technology, Institute of Quantum Matter, South China Normal University, Guangzhou 510006, China}
\address{Guangdong Basic Research Center of Excellence for Structure and Fundamental Interactions of Matter, Guangdong Provincial Key Laboratory of Nuclear Science, Guangzhou 510006, China}

\author{Hui Zhang}
\email{Mr.zhanghui@m.scnu.edu.cn}
\address{State Key Laboratory of Nuclear Physics and Technology, Institute of Quantum Matter, South China Normal University, Guangzhou 510006, China}
\address{Guangdong Basic Research Center of Excellence for Structure and Fundamental Interactions of Matter, Guangdong Provincial Key Laboratory of Nuclear Science, Guangzhou 510006, China}

\begin{abstract}

We construct an exact formulation for the solution of Bjorken expanding relativistic viscous hydrodynamics with a constant relaxation time and derive its late time transseries structure.
The two fundamental modes of the linear auxiliary equation separate the hydrodynamic contribution from the exponentially suppressed transient mode. We show that this single transient mode generates the entire hierarchy of nonperturbative sectors through nonlinear mode mixing, while its decay explains the loss of initial condition dependence and the emergence of late time hydrodynamic attractor behavior.
The hydrodynamic gradient expansion is factorially divergent for generic values of the transport parameter but converges at the discrete values $a=N^2$, where the auxiliary asymptotic expansions terminate. We further determine the exact Stokes constant of the auxiliary linear equation and establish its relation to the logarithmic Borel singularity coefficient. The vanishing of the Stokes constant at these exceptional values establishes a direct connection between auxiliary series termination, hydrodynamic convergence, and Stokes decoupling between the hydrodynamic and the nonhydrodynamic transient sector.

\end{abstract}

\maketitle

\section{Introduction}

Relativistic viscous hydrodynamics provides a remarkably successful description of the space-time evolution of the quark-gluon plasma produced in ultrarelativistic heavy-ion collisions at RHIC and the LHC, for reviews see~\cite{Heinz:2013th,Gale:2013da}. 
Its ability to reproduce a wide range of collective observables suggests that the created matter behaves as a strongly interacting relativistic fluid with small dissipative effects. 
However, similar signatures of collective behavior have also been observed in proton-nucleus and proton-proton collisions~\cite{CMS:2012qk,CMS:2010ifv,Nagle:2018nvi}, indicating that hydrodynamics may remain applicable even in systems with short lifetimes, large gradients, and substantial pressure anisotropies~\cite{Romatschke:2017vte,Kurkela:2015qoa}. 
These observations challenge the conventional interpretation of hydrodynamics as a long-wavelength effective theory restricted to systems close to local equilibrium. 
Studies based on viscous and anisotropic hydrodynamics, kinetic theory, and holography instead indicate that hydrodynamic evolution can emerge far from equilibrium, before isotropization or thermalization is achieved~\cite{Heller:2011ju,Denicol:2016bjh,Heller:2016rtz,Strickland:2018ayk}. 
This apparent universality has shifted attention from the onset of local equilibrium to the mechanisms by which microscopic information is lost and macroscopic fluid behavior emerges. 
In this context, hydrodynamic attractors and the resurgent structure of the gradient expansion provide complementary frameworks for characterizing the transition from transient nonhydrodynamic dynamics to universal hydrodynamic evolution, see~\cite{Florkowski:2017olj,Romatschke:2017ejr,Shen:2020mgh} for reviews.

The hydrodynamic attractor shows that hydrodynamization need not require convergence of the gradient expansion or complete local equilibration. 
Instead, transient nonhydrodynamic modes decay, causing solutions with different initial conditions to approach a universal trajectory~\cite{Heller:2015dha}. 
Such attractors have been identified in kinetic theory, viscous and anisotropic hydrodynamics, and strongly coupled holographic systems~\cite{Heller:2016rtz,Denicol:2016bjh,Romatschke:2017vte,Strickland:2018ayk}. 
The hydrodynamic attractor admits a fixed point interpretation as a special trajectory connecting the early time collisionless fixed point to the late time hydrodynamic fixed point~\cite{Blaizot:2017ucy,Blaizot:2019scw,Blaizot:2021cdv}.
Recent developments have related attractor behavior to dynamical dimensionality reduction in phase space and, in kinetic theory, to adiabatic hydrodynamization governed by slowly evolving pre-hydrodynamic modes~\cite{Heller:2020anv,Spalinski:2025ngd,Brewer:2019oha,Rajagopal:2024lou}.
Resurgence further shows that the divergent hydrodynamic gradient expansion encodes information about transient nonhydrodynamic sectors, with transseries relating the hydrodynamic and nonhydrodynamic contributions~\cite{Heller:2015dha,Basar:2015ava,Aniceto:2015mto,Heller:2016rtz,Aniceto:2018uik}.
This framework has since been explored beyond conformal boost-invariant Bjorken flow, including nonconformal and spatially inhomogeneous systems, alternative expansion geometries such as Gubser and Hubble flow, realistic Yang--Mills kinetic theory, and fluctuating hydrodynamics~\cite{Romatschke:2017acs,Chattopadhyay:2021ive,Behtash:2017wqg,Behtash:2018moe,Dash:2020zqx,Chen:2024pez,Chen:2023vrk,Chen:2022ryi,Du:2021fok,Du:2022bel,Kamata:2022ola,Aniceto:2022dnm,
Heller:2020jif,Behtash:2019qtk,Denicol:2018pak}, for review see~\cite{Aniceto:2018bis,Soloviev:2021lhs,Jankowski:2023fdz}.

An important step toward the explicit analytical characterization of hydrodynamic attractors was made by Denicol and Noronha, who obtained an analytical expression for the attractor in terms of special functions for Bjorken flow with a constant shear relaxation time~\cite{Denicol:2017lxn}.
In this case, for a special parameter choice, the perturbative hydrodynamic sector reduces to a convergent geometric series whose ordinary sum coincides with the exact hydrodynamic branch. 
This example provides a simple realization of a direct relation between the hydrodynamic expansion and the attractor solution.
The hydrodynamic gradient expansion of the Israel–Stewart-like hard sphere model converges absolutely within a finite range of the Knudsen number~\cite{Denicol:2019lio}.
Related analytical studies exploiting similar linearizable structures in Bjorken flow attractor equations were carried out in Refs.~\cite{Jaiswal:2019cju,Blaizot:2020gql}, while an alternative analytic approach was discussed in Ref.~\cite{Aniceto:2024pyc}.
A more radical approach is to derive the nonlinear attractor equation directly from an underlying linear system.
In particular, Blaizot and Yan showed that a coupled linear two mode system can be recast as a single nonlinear attractor equation~\cite{Blaizot:2021cdv}, while the corresponding analytical solution was used to investigate the resurgent structure of the hydrodynamic expansion~\cite{Blaizot:2020gql}.
These analytical approaches reveal that the nonlinear attractor dynamics can often be traced back to an underlying linear structure. 
This provides a unifying perspective on the emergence of closed form attractor solutions in otherwise nonlinear far from equilibrium dynamics.

The mathematical framework connecting transseries, Borel summation, and Stokes phenomena in differential equations has been systematically developed in Refs.~\cite{Costin1998,costin2006,LodayRichaud2011}. The relation between Stokes phenomena in second order linear equations and their associated Riccati systems has also been investigated in Refs.~\cite{Lambert2008,OldeDaalhuis2005}, providing a mathematical setting for studying the connection between attractive and repulsive asymptotic branches.
There exist a direct link between the transseries structure of the nonlinear Riccati system and the Stokes data of the underlying linear equation. 
For parameter-dependent second order linear differential equations, Stokes constants are analytic functions of the equation parameters under suitable normalization, and their zeros have been rigorously investigated \cite{Sibuya1975,OldeDaalhuisOlver:1994,Trinh2006}. As a parameter approaches such a zero, the corresponding Stokes constant tends to zero, leading to the disappearance of Stokes-induced coupling between the associated asymptotic branches across the relevant Stokes line.

Recent studies of Bjorken spin hydrodynamics have identified attractor behavior in nonlinear variables constructed from two linear spin density modes~\cite{Zhou:2026wjx}. 
This projective formulation relates the two dimensional linear asymptotic solution space to a one-parameter family of nonlinear solutions and provides a direct connection between the underlying linear dynamics and the corresponding attractor structure.
Motivated by these developments, in the present work we provide a systematic classification of the exceptional convergent cases considered in Ref.~\cite{Denicol:2017lxn} in terms of the transport parameters. 
We will express the nonlinear evolution equation in terms of the two fundamental late time asymptotic solutions of an auxiliary linear problem.
Within this framework, the perturbative hydrodynamic branch, the exponentially suppressed transient contribution, and its nonlinear descendants are generated from a single mode ratio, while the previously known convergent cases are unified by the termination of the auxiliary asymptotic expansions at $a=N^2$.
We further determine the exact Stokes constant and establish its relation to the logarithmic Borel singularity coefficient. The vanishing of this constant at $a=N^2$ connects auxiliary-series termination, hydrodynamic convergence, and Stokes decoupling.

The paper is organized as follows.
In Sec.~\ref{sec:convergent-gradient-attractor}, we review the special
convergent attractor solution found in Ref.~\cite{Denicol:2017lxn}.
In Sec.~\ref{sec:Exact_Stokes_Constant}, we analytically determine the Stokes constant from the exact solution, providing a benchmark for comparison with the asymptotic projection formulation developed in the next section.
Using the late time asymptotic solution, we reformulate the late time transseries in
Sec.~\ref{sec:transseries-attractor}.
We then investigate the dependence on the transport parameters,
the convergence properties of the hydrodynamic expansion, and
the causality constraints.
The large order behavior is analyzed through the recurrence relations
of the auxiliary second order linear equation and the Borel singularity
and resurgence structure of its asymptotic solutions.
We also present numerical large order diagnostics of the expansion
coefficients.
Finally, Sec.~\ref{sec:summary} presents the discussion and conclusions.

\section{Review of a Convergent Gradient Expansion in Denicol-Noronha Model}
\label{sec:convergent-gradient-attractor}

An instructive example in which the perturbative hydrodynamic sector can be summed directly to the exact hydrodynamic branch was discussed by Denicol and Noronha in Ref.~\cite{Denicol:2017lxn}. 
In this work, we focus exclusively on the late time hydrodynamic branch, without imposing early time regularity conditions.

For a Bjorken expanding fluid with a constant relaxation time and $\lambda=0$, the evolution equation for the normalized shear stress $\chi(\hat{\tau})$ reduces to the Riccati-type evolution equation
\begin{equation}
    \hat{\tau}\frac{d\chi}{d\hat{\tau}}
    +\frac{4}{3}\chi^{2}
    +\hat{\tau}\chi
    -\frac{3a}{4}
    =0,
    \label{eq:denicol-riccati}
\end{equation}
where $\hat{\tau}$ denotes the proper time measured in units of the relaxation time, while the dimensionless coefficient $a$ is determined by the transport coefficients of the underlying hydrodynamic theory.

The perturbative, or hydrodynamic, sector is defined by the formal late time gradient expansion
\begin{equation}
    \chi^{(0)}(\hat{\tau})
    =
    \frac{3a}{4}
    \sum_{n=0}^{\infty}
    \frac{c_n}{\hat{\tau}^{n}},
    \qquad
    c_0=0,
    \qquad
    c_1=1.
    \label{eq:denicol-gradient-expansion}
\end{equation}
Substituting Eq.~\eqref{eq:denicol-gradient-expansion} into Eq.~\eqref{eq:denicol-riccati} yields the nonlinear recursion relation
\begin{equation}
    c_{n+1}
    =
    n c_n
    -
    a\sum_{m=0}^{n}c_{n-m}c_m, 
    \qquad
    n \geq 1.
    \label{eq:denicol-recursion}
\end{equation}
For parameter values compatible with the standard linear causality and stability condition, $a\leq 8/9$, the first term on the right-hand side of Eq.~\eqref{eq:denicol-recursion} produces factorial growth at large order. Consequently, the corresponding gradient expansion is generically asymptotic rather than convergent.

A qualitatively different behavior occurs for the special choice
\begin{equation}
    a=1 > \frac{8}{9}.
    \label{eq:denicol-a-one}
\end{equation}
Besides \(a=1\), the authors of Ref.~\cite{Denicol:2017lxn} also note that \(a=4\) and \(a=9\) yield convergent gradient expansions, although these values lie outside the causal and stable regime.
For \(a=1\), Eq.~\eqref{eq:denicol-recursion} implies
\begin{equation}
    c_n=1,
    \qquad
    n\geq 1.
    \label{eq:denicol-coefficients-one}
\end{equation}
The hydrodynamic sector therefore reduces to the geometric series
\begin{equation}
    \chi^{(0)}(\hat{\tau})
    =
    \frac{3}{4}
    \sum_{n=1}^{\infty}
    \frac{1}{\hat{\tau}^{n}}.
    \label{eq:denicol-geometric-series}
\end{equation}
For $\hat{\tau}>1$, the series converges and can be summed explicitly,
\begin{equation}
    \chi^{(0)}(\hat{\tau})
    =
    \frac{3}{4}
    \frac{\hat{\tau}^{-1}}{1-\hat{\tau}^{-1}}
    =
    \frac{3}{4(\hat{\tau}-1)}.
    \label{eq:denicol-resummed-sector}
\end{equation}
The dimensionless proper time satisfies $\hat{\tau}>0$ in general.
For the special case $a=1$, however, the convergent gradient representation and the corresponding hydrodynamic branch are considered on the late time domain $\hat{\tau}>1$, since the solution has a pole at $\hat{\tau}=1$.

The significance of Eq.~\eqref{eq:denicol-resummed-sector} becomes apparent upon comparison with the full analytic solution. 
For $a=1$, the general solution of Eq.~\eqref{eq:denicol-riccati} can be written as
\begin{equation}
    \left.
    \chi(\hat{\tau};\sigma)
    \right|_{a=1}
    =
    \frac{3}{4}
    \frac{
        \displaystyle
        \hat{\tau}^{-1}
        -
        \sigma
        \left(
            1+\hat{\tau}^{-1}
        \right)
        e^{-\hat{\tau}}
    }{
        \displaystyle
        1-\hat{\tau}^{-1}
        +
        \sigma\hat{\tau}^{-1}e^{-\hat{\tau}}
    },
    \label{eq:denicol-full-solution-a-one}
\end{equation}
where $\sigma$ parametrizes the initial condition.  
Here we have corrected a sign typo in Eq.~(24) of Ref.~\cite{Denicol:2017lxn}.
The hydrodynamic branch is obtained by setting $\sigma=0$, which removes the transient contribution and gives
\begin{equation}
    \left.
    \chi_{\mathrm{hyd}}(\hat{\tau})
    \right|_{a=1}
    =
    \frac{3}{4(\hat{\tau}-1)}.
    \label{eq:denicol-attractor-a-one}
\end{equation}
Equations~\eqref{eq:denicol-resummed-sector} and \eqref{eq:denicol-attractor-a-one} therefore imply
\begin{equation}
    \left.
    \chi^{(0)}(\hat{\tau})
    \right|_{a=1}
    =
    \left.
    \chi_{\mathrm{hyd}}(\hat{\tau})
    \right|_{a=1},
    \qquad
    \hat{\tau}>1.
    \label{eq:denicol-sector-equals-attractor}
\end{equation}
Thus, in this particular example of Ref.~\cite{Denicol:2017lxn}, the convergent perturbative sector sums directly to the exact hydrodynamic branch.

The separation between the hydrodynamic branch and the transient dependence on the initial condition can be made explicit by expanding Eq.~\eqref{eq:denicol-full-solution-a-one} at late times
\begin{equation}
     \left.
    \chi(\hat{\tau};\sigma)
    \right|_{a=1}
    =
    \frac{3}{4(\hat{\tau}-1)}
    -
    \frac{3}{4}
    \sum_{n=1}^{\infty}
    (-1)^{n-1}
    \sigma^{n}
    e^{-n\hat{\tau}}
    \frac{\hat{\tau}^{2}}
         {(\hat{\tau}-1)^{n+1}}.
    \label{eq:denicol-transseries-a-one}
\end{equation}
The first term is the hydrodynamic sector, whereas the exponentially suppressed terms encode the dependence on the initial condition. In this representation, $\sigma$ plays the role of a transseries parameter.

It is important to emphasize that the choice $a=1$ lies outside the causal and linearly stable domain $a\leq 8/9$.
The equality between the ordinarily summed perturbative sector and the exact hydrodynamic branch should therefore not be interpreted as a generic property of physically admissible relativistic hydrodynamic theories.
In the causal regime, the gradient expansion is generally divergent, and reconstructing the full physical attractor requires an appropriate resummation prescription together with a consistent treatment of the exponentially suppressed nonhydrodynamic sectors and the relevant boundary conditions.

\section{Exact Stokes Constant in Denicol-Noronha Model}
\label{sec:Exact_Stokes_Constant}

After reviewing the Denicol–Noronha model in the previous section, we now directly compute the Stokes constant using the exact analytical solution obtained in Ref.~\cite{Denicol:2017lxn}.
This calculation does not rely on the large order relations.

Using Eq.~(14) of Ref.~\cite{Denicol:2017lxn}, the two fundamental solutions of the auxiliary linear equation associated with Eq.~\eqref{eq:denicol-riccati} can be written as
\begin{equation}
\begin{aligned}
    y_M(\hat{\tau})
    &=
    \hat{\tau}^{-1/2}e^{-\hat{\tau}/2}
    M_{-1/2,\nu}(\hat{\tau}),
    \qquad
    y_W(\hat{\tau})
    &=
    \hat{\tau}^{-1/2}e^{-\hat{\tau}/2}
    W_{-1/2,\nu}(\hat{\tau}),
\end{aligned}
\label{eq:stokes-exact-fundamental}
\end{equation}
where we denote $\nu=\sqrt{a}$. Their late time behavior is
\begin{equation}
\begin{aligned}
    y_M(\hat{\tau})
    &\sim
    \frac{\Gamma(1+2\nu)}{\Gamma(1+\nu)}
    \left[1+O(\hat{\tau}^{-1})\right],
    \qquad
    y_W(\hat{\tau})
    &\sim
    \frac{e^{-\hat{\tau}}}{\hat{\tau}}
    \left[1+O(\hat{\tau}^{-1})\right].
\end{aligned}
\label{eq:stokes-exact-asymptotics}
\end{equation}
These asymptotic behaviors will be used to fix the normalization of the solutions and extract the Stokes constant.

As shown in Sec.~\ref{subsubsec:Borel_Singularities}, the formal asymptotic solution $\Phi_+$ of the auxiliary linear equation has a Borel singularity at $\xi=1$. 
We denote its two lateral Borel resummations by $y_+^{(\pm)}(\hat{\tau})=\mathcal{S}_{0^\pm}\Phi_+(\hat{\tau})$. 
Using the Laplace integral representation of Tricomi's confluent hypergeometric function, these solutions can be expressed as
\begin{equation}
\begin{aligned}
    y_+^{(\pm)}(\hat{\tau})
    &=
    e^{\pm i\pi\nu}\hat{\tau}^{\nu}
    U\left(
        \nu,1+2\nu,
        \hat{\tau}e^{\pm i\pi}
    \right)
    =
    \frac{\Gamma(1+\nu)}{\Gamma(1+2\nu)}
    y_M(\hat{\tau})
    +
    \nu e^{\mp i\pi\nu}y_W(\hat{\tau}),
\end{aligned}
\label{eq:stokes-exact-connection}
\end{equation}
where the second equality follows from the connection formulas of the confluent hypergeometric functions.

Taking the difference between the two lateral
solutions gives the exact relation
\begin{equation}
\begin{aligned}
    y_+^{(+)}-y_+^{(-)}
    &=
    \nu\left(
        e^{-i\pi\nu}-e^{i\pi\nu}
    \right)y_W
    =
    -2i\nu\sin(\pi\nu)y_W.
\end{aligned}
\label{eq:stokes-exact-jump}
\end{equation}
Using the normalization of $y_W$ given in Eq.~\eqref{eq:stokes-exact-asymptotics}, the Stokes constant of the auxiliary linear equation is
\begin{equation}
    S_y(a)=-2i\nu\sin(\pi\nu).
    \label{eq:stokes-exact-linear-result}
\end{equation}
These results, obtained directly from the exact special function solutions, provide an independent check of the large order analysis presented in Sec.~\ref{subsubsec:Borel_Singularities}.

\section{Projective Formulation of Late Time Transseries}
\label{sec:transseries-attractor}

Denicol and Noronha obtained an exact solution by linearizing the nonlinear equation~\cite{Denicol:2017lxn}. Here we first construct the late time transseries directly from the auxiliary linear equation, following Ref.~\cite{Zhou:2026wjx}. 
While Ref.~\cite{Zhou:2026wjx} used the projective structure to study the Stokes connection between the attractor and repellor, we employ it to distinguish the hydrodynamic branch from nonhydrodynamic transient contributions and to establish the relation between the convergence of the gradient expansion and Stokes phenomena.
Unlike the spin density case, the linear equation is written in terms of an auxiliary variable without a direct physical interpretation.

Introducing $y(\hat{\tau})$ through
\begin{equation}
    \chi(\hat{\tau})
    =
    \frac{3\hat{\tau}}{4}
    \frac{y'(\hat{\tau})}{y(\hat{\tau})},
    \label{eq:chi-log-derivative}
\end{equation}
we obtain the second order linear equation
\begin{equation}
    y''(\hat{\tau})
    +
    \left(1+\frac{1}{\hat{\tau}}\right)y'(\hat{\tau})
    -\frac{a}{\hat{\tau}^{2}}y(\hat{\tau})
    =0.
    \label{eq:linearized-y-equation}
\end{equation}

Following Ref.~\cite{Zhou:2026wjx}, we seek two linearly independent formal late time solutions of Eq.~\eqref{eq:linearized-y-equation} in the form
\begin{equation}
    y_{\pm}(\hat{\tau})
    =e^{-A_{\pm}\hat{\tau}}\hat{\tau}^{\beta_{\pm}}
    \Phi_{\pm}(\hat{\tau}),
    \label{eq:y-transseries-ansatz}
    \qquad
    \Phi_{\pm}(\hat{\tau})=
    \sum_{n=0}^{\infty}\frac{c_n^{(\pm)}}{\hat{\tau}^{n}},
    \qquad c_0^{(\pm)}\neq0.
\end{equation}
Equalities involving inverse-power series are understood in the formal sense. Substitution gives $A_+=0$, $A_-=1$, $\beta_+=0$, and $\beta_-=-1$, so that
\begin{align}
    y_+(\hat{\tau})
    &=
    \Phi_+(\hat{\tau})
    =\sum_{n=0}^{\infty}\frac{c_n^{(+)}}{\hat{\tau}^{n}},
    \label{eq:y-plus-sector}\\
    y_-(\hat{\tau})
    &=
    e^{-\hat{\tau}}\hat{\tau}^{-1}\Phi_-(\hat{\tau})
    =e^{-\hat{\tau}}\hat{\tau}^{-1}
    \sum_{n=0}^{\infty}\frac{c_n^{(-)}}{\hat{\tau}^{n}}.
    \label{eq:y-minus-sector}
\end{align}
The coefficients satisfy
\begin{equation}
    c_{n+1}^{(+)}=\frac{n^2-a}{n+1}c_n^{(+)},
    \label{eq:plus-recursion}
    \qquad
    c_{n+1}^{(-)}=-\frac{(n+1)^2-a}{n+1}c_n^{(-)},
    \qquad n\geq0.
\end{equation}
The algebraic branch $y_+$ determines the hydrodynamic sector through Eq.~\eqref{eq:chi-log-derivative}, while $y_-$ provides the exponentially suppressed contribution.

\subsubsection{Representation of the Transseries}
\label{subsec:mode-ratio-reparametrization}

At the level of formal late time expansions, we write
$y=C_+y_++C_-y_-$. 
Since Eq.~\eqref{eq:chi-log-derivative} is invariant under an overall rescaling of $y$, the nonlinear solution depends only on $\sigma=C_-/C_+$ and can be written as
\begin{equation}
    \chi(\hat{\tau};\sigma)
    =\frac{3\hat{\tau}}{4}
    \frac{d}{d\hat{\tau}}
    \ln\left[y_+(\hat{\tau})+\sigma y_-(\hat{\tau})\right].
    \label{eq:chi-general-transseries}
\end{equation}
We define the mode ratio and the corresponding nonlinear branches by
\begin{equation}
    R(\hat{\tau})\equiv\frac{y_-(\hat{\tau})}{y_+(\hat{\tau})},
    \qquad
    \chi_{\pm}(\hat{\tau})\equiv
    \frac{3\hat{\tau}}{4}\frac{y_{\pm}'(\hat{\tau})}{y_{\pm}(\hat{\tau})}.
    \label{eq:mode-ratio-definition}
\end{equation}
Since $R(\hat{\tau})\sim e^{-\hat{\tau}}\hat{\tau}^{-1}$, for finite $\sigma$ one finds at late times
\begin{equation}
    \chi(\hat{\tau};\sigma)-\chi_+(\hat{\tau})
    =\frac{3\hat{\tau}}{4}
    \frac{\sigma R'(\hat{\tau})}{1+\sigma R(\hat{\tau})}
    =\mathcal O(e^{-\hat{\tau}}).
\end{equation}
Thus, regular solutions with finite $\sigma$ approach the late time hydrodynamic branch $\chi_+$. 
The exponentially suppressed auxiliary solution $y_-$ is associated with the nonhydrodynamic transient sector through the mode ratio $R=y_-/y_+$, while its logarithmic derivative defines the repeller branch $\chi_-$. The $n=1$ transseries sector provides the leading nonhydrodynamic correction, while higher sectors arise from nonlinear powers of the same mode ratio.

To obtain the full transseries, we write
\begin{equation}
    R(\hat{\tau})
    =e^{-\hat{\tau}}\hat{\tau}^{-1}\Psi(\hat{\tau}),
    \qquad
    \Psi(\hat{\tau})\equiv
    \frac{\Phi_-(\hat{\tau})}{\Phi_+(\hat{\tau})}.
    \label{eq:R-Psi-decomposition}
\end{equation}
The ratio $\Psi=\sum_{k=0}^{\infty}\psi_k\hat{\tau}^{-k}$ is understood formally, with $\psi_0=1$. Using $R'/R=-1-1/\hat{\tau}+\Psi'/\Psi$, expansion of Eq.~\eqref{eq:chi-general-transseries} in powers of $\sigma R$ gives the formal transseries
\begin{equation}
    \chi(\hat{\tau};\sigma)
    =\chi^{(0)}(\hat{\tau})
    +\sum_{n=1}^{\infty}
    \sigma^n e^{-n\hat{\tau}}\hat{\tau}^{1-n}
    \mathcal F_n(\hat{\tau}),
    \qquad
    \chi^{(0)}(\hat{\tau})\equiv\chi_+(\hat{\tau}),
    \label{eq:chi-full-transseries-Psi}
\end{equation}
where
\begin{equation}
    \mathcal F_n(\hat{\tau})
    =(-1)^{n-1}\frac34
    \left[\Psi^{n-1}(\hat{\tau})\Psi'(\hat{\tau})
    -\left(1+\frac1{\hat{\tau}}\right)\Psi^n(\hat{\tau})\right]
    =\sum_{k=0}^{\infty}\frac{f_{n,k}}{\hat{\tau}^k}.
    \label{eq:Fn-definition}
\end{equation}
The $n=1$ sector is the leading transient correction. Higher sectors arise from nonlinear powers of the same mode ratio and do not correspond to additional independent modes of the auxiliary linear equation.

\subsubsection{Equivalence to the Standard Transseries Ansatz}
\label{sec:equivalent-transseries}

The standard one-parameter transseries ansatz in relativistic hydrodynamics~\cite{Heller:2015dha,Basar:2015ava,Aniceto:2015mto} is
\begin{equation}
    \chi(\hat{\tau};\sigma)
    =\sum_{n=0}^{\infty}
    \sigma^n e^{-nA\hat{\tau}}\hat{\tau}^{nB}\Phi_n(\hat{\tau}),
    \qquad
    \Phi_n(\hat{\tau})
    =\sum_{k=0}^{\infty}\frac{c_{n,k}}{\hat{\tau}^k}.
\end{equation}
Here $A$ and $B$ determine the exponential and algebraic weights. The equalities involving formal series have the same meaning as above. For the present system, $A=1$ and $B=0$, since the leading correction $\hat{\tau}R'$ is proportional to $e^{-\hat{\tau}}$. Thus,
\begin{equation}
    \chi(\hat{\tau};\sigma)
    =
    \sum_{n=0}^{\infty}\sigma^n e^{-n\hat{\tau}}\Phi_n(\hat{\tau}).
    \label{eq:standard-transseries}
\end{equation}
The perturbative coefficients are related to those in Eq.~\eqref{eq:denicol-gradient-expansion} by $c_{0,k}(a)=\frac{3a}{4}c_k(a)$. Comparing with Eq.~\eqref{eq:chi-full-transseries-Psi}, we obtain
\begin{equation}
    \Phi_n(\hat{\tau})=\hat{\tau}^{1-n}\mathcal F_n(\hat{\tau}),
    \qquad
    f_{n,k}=c_{n,k+n-1},
    \qquad n\geq1,\quad k\geq0.
    \label{eq:coefficient-shift}
\end{equation}
Consequently, $c_{n,0}=\cdots=c_{n,n-2}=0$ for $n\geq2$. The two representations differ only in whether the factor $\hat{\tau}^{1-n}$ is displayed explicitly or absorbed into $\Phi_n$. This factor arises from the logarithmic-derivative transformation and should be retained in trans-asymptotic reorganizations.

In summary, the algebraic auxiliary solution $y_+$ determines the hydrodynamic branch $\chi_+$, which corresponds to the zeroth sector of the nonlinear transseries. The exponentially suppressed auxiliary solution $y_-$ determines the repeller branch $\chi_-$ through the logarithmic-derivative transformation. The nonhydrodynamic transient sectors are generated by the mode ratio $R=y_-/y_+$, with the leading transient contribution and its nonlinear descendants arising from the same auxiliary solution.

\section{Large-order Behavior and Stokes Phenomena}
\label{subsec:a-causality-large order}

In the following, we first examine the dependence of the transseries sectors
and their large order behavior on the transport parameter $a$. We then trace
the origin of divergence and convergence to the auxiliary linear problem and
finally relate the generic factorial growth to the Stokes coupling between the
hydrodynamic and transient branches.
In the Denicol--Noronha model~\cite{Denicol:2017lxn}, the dimensionless parameter $a$ controls the relative magnitude of the shear viscous timescale and the shear relaxation time.
The standard linearized causality and stability conditions constrain this dimensionless parameter as~\cite{Denicol:2017lxn,Rocha:2023ilf}
\begin{equation}
0 \leq a \leq \frac{8}{9}.
\label{eq:a-causality-bound}
\end{equation}
The lower bound \(a=0\) corresponds to the vanishing-viscosity limit,  while the upper bound
$a=8/9$ is specific to the relaxation-type theory considered here.
In the following, we consider the general case \(a>0\).

\subsubsection{Transport Parameter Dependence and large order Diagnostics}
\label{subsubsec:large_order_Diagnostics}

To make the role of the transport parameter $a$ explicit, we now examine how it enters the asymptotic structure of the nonlinear transseries sectors. The nonlinear sectors inherit their dependence on the transport parameter $a$ through $\Psi(\hat{\tau})$ in Eqs.~\eqref{eq:R-Psi-decomposition} and \eqref{eq:Fn-definition}.
For example, the late time expansion of $\Psi(\hat{\tau})$ explicitly shows the dependence on the transport parameter $a$,
\begin{equation}
    \Psi(\hat{\tau})
    =
    1
    +
    \frac{2a-1}{\hat{\tau}}
    +
    \mathcal{O}\!\left(\hat{\tau}^{-2}\right).
    \label{eq:Psi-first-a-coefficient}
\end{equation}
Accordingly,
\begin{equation}
    \chi(\hat{\tau};\sigma,a)
    =
    \chi_{+}(\hat{\tau};a)
    +
    \sum_{n=1}^{\infty}
    \sigma^n
    e^{-n\hat{\tau}}
    \hat{\tau}^{1-n}
    \sum_{k=0}^{\infty}
    \frac{f_{n,k}(a)}
         {\hat{\tau}^{k}},
    \label{eq:chi-transseries-explicit-a}
\end{equation}
where the first two coefficients are
\begin{align}
    f_{n,0}(a)
    &=
    (-1)^n\frac{3}{4},
    \nonumber\\
    f_{n,1}(a)
    &=
    (-1)^n\frac{3}{4}
    \left[
        n(2a-1)
        +
        1
    \right].
    \label{eq:f-first-a-coefficients}
\end{align}
Thus, $a$ enters the perturbative sector already at its leading nonvanishing order as seen from Eq.(\ref{eq:denicol-gradient-expansion}), whereas it first appears at subleading order in each nonperturbative sector.

\begin{figure}[h]
    \centering
    \includegraphics[width=0.55\linewidth]
    {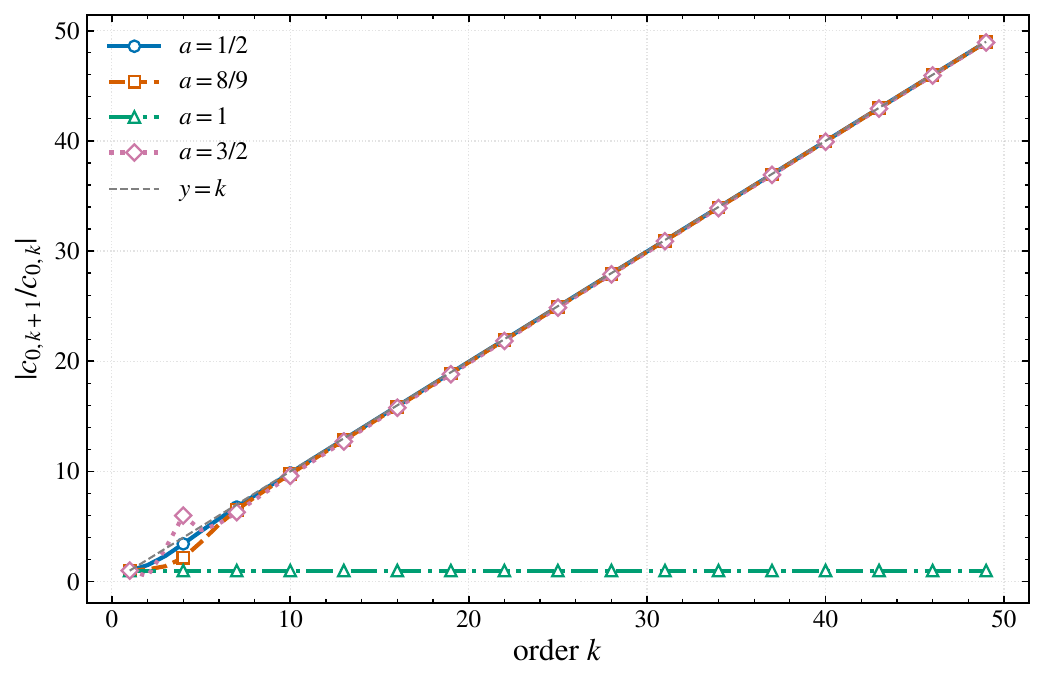}
    \caption{Ratios of successive hydrodynamic coefficients $|c_{0,k+1}/c_{0,k}|$ for selected values of $a$.}
    \label{fig:a-hydrodynamic-large-order}
\end{figure}

To analyze the large order behavior, we use the perturbative coefficients introduced in Eq.~\eqref{eq:denicol-gradient-expansion}, together with their recurrence relation in Eq.~\eqref{eq:denicol-recursion}.
For convenience, we denote the perturbative coefficients by $c_{0,k}(a)$.
In Fig.~\ref{fig:a-hydrodynamic-large-order}, we compare the selected values $a=1/2$, $8/9$, $1$, and $3/2$. The values $a=1/2$, $8/9$, and $3/2$ display factorial growth, whereas $a=1$ yields the convergent geometric expansion in Eq.~\eqref{eq:denicol-attractor-a-one}. 
The large-$k$ behavior of the successive coefficient ratios provides a numerical diagnostic of factorial divergence and finite-radius convergence.
This numerical comparison is consistent with the analytic large order behavior discussed below and indicates that crossing the causal boundary does not by itself change the large order structure.

\subsubsection{Large Order Behavior from the Linear Recurrence}
\label{subsubsec:Linear_Recurrence}

We now show analytically that the contrasting large order behaviors observed above originate from the recurrence structure of the two fundamental solutions of the underlying linear problem.
The recurrence relation in Eq.~\eqref{eq:plus-recursion} for the coefficients of the projected auxiliary solution $\Phi_+$ gives
\begin{equation}
    \frac{c_{n+1}^{(+)}}{c_n^{(+)}}
    =
    \frac{n^2-a}{n+1}
    =
    n-1+\mathcal{O}\!\left(n^{-1}\right),
    \qquad
    n\rightarrow\infty.
\end{equation}
For generic values of $a$, we have
\begin{equation}
    c_n^{(+)}
    \sim
    C\,\Gamma(n-1),
\end{equation}
where $\Gamma$ denotes the Euler gamma function. 
The inverse-power expansion of the auxiliary solution is therefore factorially divergent.
Moreover, the recurrence in Eq.~\eqref{eq:plus-recursion} can be solved explicitly. 
Since the recurrence coefficient can be factorized, its iteration can be expressed as a product and
then written in terms of Gamma functions. 
One obtains
\begin{equation}
    c_n^{(+)}
    =
    \frac{
        \Gamma(n-\nu)\Gamma(n+\nu)
    }{
        \Gamma(-\nu)\Gamma(\nu)\Gamma(n+1)
    },
    \label{eq:plus-coefficients-gamma}
\end{equation}
where $\nu\equiv\sqrt{a}$, and the expression applies to generic $\nu\notin\mathbb{Z}$.

The large order expansion of $c_n^{(+)}$ can be written as
\begin{equation}
    c_n^{(+)}
    \sim
    -\frac{\nu\sin(\pi \nu)}{\pi}
    \Gamma(n-1)
    \left[
        1+\frac{a-1}{n}
        +\mathcal{O}\!\left(n^{-2}\right)
    \right],
    \label{eq:plus-coefficients-large-order-explicit}
\end{equation}
where we have used the reflection formula.

This linear large order behavior can be related directly to the coefficients
of the hydrodynamic sector. Using the expansion of $\Phi_+$ in
Eq.~\eqref{eq:y-plus-sector}, together with the hydrodynamic sector
$\Phi_0$ defined in Eq.~\eqref{eq:standard-transseries} and the
logarithmic-derivative relation in Eq.~\eqref{eq:chi-log-derivative},
matching the inverse-power coefficients gives
\begin{equation}
    c_{0,n}
    =
    -\frac{3n}{4}c_n^{(+)}
    -
    \sum_{m=1}^{n-1}
    c_m^{(+)}c_{0,n-m}.
    \label{eq:linear-to-hydro-coefficients}
\end{equation}
At large order, the first term gives the leading factorial behavior, while
the convolution contributes at subleading order. This gives
\begin{equation}
    c_{0,n}
    \sim
    -\frac{3n}{4}c_n^{(+)}
    \sim
    \frac{3\nu \sin(\pi\nu )}{4\pi}\Gamma(n),
\end{equation}
and therefore
\begin{equation}
    \frac{c_{0,n+1}}{c_{0,n}}
    \sim
    n.
\end{equation}
The factorial growth of the hydrodynamic coefficients follows directly from the large order behavior of the auxiliary linear solution, consistent with the result obtained from the nonlinear Riccati recursion.
The underlying origin of this divergence is further explored in the next subsection through the Stokes structure of the linear problem.

A different behavior occurs at the special values
\begin{equation}
    a=N^2,\qquad N\in\mathbb{N}.
\end{equation}
In this case, the recurrence relation terminates at $n=N$, so that $c_{N+1}^{(+)}=0$ and all higher coefficients vanish. 
As a result, $\Phi_+$ becomes a finite polynomial in $\hat{\tau}^{-1}$, and its logarithmic derivative $\chi_+$ is a rational function. 
The corresponding inverse-power expansion therefore has a nonzero radius of convergence determined by the nearest pole.
This behavior is also reflected in the large order amplitude. 
As
$\nu\to N$, we have
\begin{equation}
    -\frac{\nu\sin(\pi\nu)}{\pi}\longrightarrow 0.
\end{equation}
The vanishing of this amplitude is consistent with the termination of the linear recurrence and the convergence of the hydrodynamic expansion at $a=N^2$.
Similarly, the recurrence relation for the auxiliary solution \(y_-\), associated with the repellor branch, also terminates, with \(c_N^{(-)}=0\). Thus, both auxiliary asymptotic expansions reduce to finite inverse-power series.

\subsubsection{Large Order Relations from Borel Singularities}
\label{subsubsec:Borel_Singularities}

The factorial divergence discussed above can also be understood from the Borel structure of the auxiliary linear equation~\eqref{eq:linearized-y-equation}. 
This analysis allows us to establish the large order relation between the two fundamental solutions and examine the convergence properties without explicitly solving the coefficient recurrence.
The point $\hat{\tau}=\infty$ is a rank-one irregular singular point of this equation. 
For such equations, Borel singularities are related to the large order behavior of formal asymptotic solutions~\cite{Olver:1997,OldeDaalhuisOlver:1994}.
This resurgent coupling relates the formal asymptotic expansions of the two branches and does not imply any linear dependence between the fundamental solutions $y_+$ and $y_-$.

Since we consider the late time regime \(\hat{\tau}\to+\infty\), we focus on the positive Borel direction, which is associated with the dominant asymptotic branch \(\Phi_+\).
For the formal inverse-power expansion of $\Phi_+(\hat{\tau})$ in Eq.~\eqref{eq:y-plus-sector}, we define the Borel transform~\cite{Olver:1997,OldeDaalhuisOlver:1994},
\begin{equation}
    \mathcal{B}\Phi_+(\xi)
    =
    \sum_{n=0}^{\infty}
    \frac{c_n^{(+)}}{n!}\xi^n.
    \label{eq:borel-plus-definition}
\end{equation}
Using Eq.~\eqref{eq:linearized-y-equation} and the Borel transform defined in Eq.~\eqref{eq:borel-plus-definition}, we obtain
\begin{equation}
    \xi(1-\xi)\frac{d^2\mathcal{B}\Phi_+}{d\xi^2}
    +(1-\xi)\frac{d\mathcal{B}\Phi_+}{d\xi}
    +a\mathcal{B}\Phi_+=0.
    \label{eq:borel-plus-equation}
\end{equation}
This hypergeometric differential equation has three regular singular points at $\xi=0$, $\xi=1$, and $\xi=\infty$. 
We first consider the solution analytic at $\xi=0$, which corresponds to the Borel transform of the late time asymptotic expansion,
\begin{equation}
    \mathcal{B}\Phi_+(\xi)
    ={}_2F_1(-\nu,\nu;1;\xi),
    \qquad \nu=\sqrt{a},
    \label{eq:borel-plus-hypergeometric}
\end{equation}
where $_2F_1$ denotes the Gauss hypergeometric function.
For generic noninteger \(\nu\), this solution has a branch point at \(\xi=1\), which governs the large-order behavior of the coefficients \(c_n^{(+)}\).

We then examine its singular behavior near $\xi=1$ to determine the large order growth of the expansion coefficients. For generic values of $a$, $\xi=1$ is a logarithmic branch point, and the local expansion can be written as
\begin{equation}
\begin{aligned}
    \mathcal{B}\Phi_+(\xi)
    ={}&H(\xi)+\mathcal{S}_+(a)(1-\xi)\log(1-\xi)
    {}_2F_1(1-\nu,1+\nu;2;1-\xi),
    \label{eq:borel-plus-local-singularity}
\end{aligned}
\end{equation}
where $H(\xi)$ denotes the analytic part near $\xi=1$.

Substituting Eq.~\eqref{eq:borel-plus-local-singularity} into Eq.~\eqref{eq:borel-plus-equation} gives
$\mathcal{S}_+(a)=-a\mathcal{B}\Phi_+(1)$. The Gauss summation formula then yields
\begin{equation}
    \mathcal{S}_+(a)
    =\frac{1}{\Gamma(-\nu)\Gamma(\nu)}
    =-\frac{\nu\sin(\pi\nu)}{\pi},
    \label{eq:stokes-coupling-explicit}
\end{equation}
where $\mathcal{S}_+(a)$ denotes the coefficient of the logarithmic Borel singularity in the large order normalization adopted here.
$\mathcal{S}_+(a)$ determines the leading factorial growth of the perturbative coefficients and is related to the Stokes constant $S_y(a)$ of the auxiliary linear equation through the Borel--Laplace discontinuity.

Comparing Eq.~\eqref{eq:stokes-coupling-explicit} with the exact result in Eq.~\eqref{eq:stokes-exact-linear-result}, we obtain
\begin{equation}
    S_y(a)=2\pi i\,\mathcal{S}_+(a).
\end{equation}
The factor of $2\pi i$ arises from the discontinuity of the logarithmic term across the Borel branch cut at $\xi=1$.
This confirms the consistency between the exact Stokes constant and the Borel singularity analysis.

The hypergeometric function appearing in the logarithmic term of Eq.~\eqref{eq:borel-plus-local-singularity} can be expanded near $\xi=1$ as
\begin{equation}
    {}_2F_1(1-\nu,1+\nu;2;1-\xi)
    =\sum_{j=0}^{\infty}
    \frac{(-1)^j c_j^{(-)}}{(j+1)!}(1-\xi)^j.
\end{equation}
This relation shows that the singular part of the Borel transform of $\Phi_+$ is directly related to the asymptotic coefficients $c_j^{(-)}$ of the exponentially suppressed branch $y_-$.
The branch cut starting at $\xi=1$ produces a Stokes discontinuity proportional to $e^{-\hat{\tau}}\hat{\tau}^{-1}\Phi_-(\hat{\tau})$ under Borel--Laplace resummation. 
This has the same asymptotic structure as the exponentially suppressed solution $y_-$ in Eq.~\eqref{eq:y-minus-sector}.

The logarithmic singularity at $\xi=1$ therefore gives the inverse-factorial large order relation
\begin{equation}
    c_n^{(+)}
    \sim
    \mathcal{S}_+(a)
    \sum_{j=0}^{\infty}
    c_j^{(-)}\Gamma(n-1-j),
    \qquad n\rightarrow\infty,
    \label{eq:linear-large-order}
\end{equation}
where the sum is understood as an asymptotic expansion. This relation
connects the large order coefficients of the $y_+$ branch to the low-order
coefficients of the $y_-$ branch of the auxiliary linear equation, without
implying any linear dependence between the two fundamental solutions.
With the normalization adopted above, the first two terms are
\begin{equation}
    c_n^{(+)}
    \sim
    \mathcal{S}_+(a)
    \left[
        \Gamma(n-1)+(a-1)\Gamma(n-2)+\cdots
    \right].
    \label{eq:linear-large-order-first-terms}
\end{equation}
This agrees with the exact large order result in
Eq.~\eqref{eq:plus-coefficients-large-order-explicit}, including the
first subleading correction. The factorial amplitude has thus been obtained
directly from the Borel singularity, independently of the exact solution
of the linear recurrence.

The logarithmic-derivative relation in Eq.~\eqref{eq:chi-log-derivative} connects the auxiliary solution $y_+$ to the hydrodynamic expansion of $\chi_+$. 
As shown in the previous subsection~\ref{subsubsec:Linear_Recurrence}, the hydrodynamic coefficients satisfy $c_{0,n}\sim -3 n c_n^{(+)} /4$.
Thus, Eq.~\eqref{eq:linear-large-order} also establishes the factorial divergence of the hydrodynamic expansion from the Borel singularity.
More explicitly, for generic values of $a$, the leading large order behavior of the hydrodynamic coefficients is given by
\begin{equation}
    c_{0,n}
    \sim
    -\frac{3}{4}\mathcal{S}_+(a)\Gamma(n)
    =
    -\frac{3}{4}\frac{S_y(a)}{2\pi i}\Gamma(n),
    \qquad n\rightarrow\infty.
\end{equation}
This shows that the leading factorial growth of the nonlinear hydrodynamic coefficients is directly controlled by the exact Stokes constant of the auxiliary linear equation.
At the exceptional values $a=N^2$, the Stokes related coefficient vanishes,
\begin{equation}
    \mathcal{S}_+(N^2)=0,
    \qquad N\in\mathbb{N}.
\end{equation}
At these exceptional values, the hypergeometric function ${}_2F_1(-N,N;1;\xi)$ reduces to a polynomial, and the logarithmic Borel singularity at $\xi=1$ disappears.
This is consistent with the termination of the linear recurrence discussed in the previous subsection~\ref{subsubsec:Linear_Recurrence}.

From the viewpoint of resurgence analysis, the vanishing of the stokes constant $S_y(a)$ at $a=N^2$ indicates the disappearance of Stokes-induced coupling between the two auxiliary branches $y_+$ and $y_-$. Through the logarithmic-derivative transformation, $y_+$ determines the zeroth hydrodynamic sector $\chi^{(0)}$, while $y_-$ generates the exponentially suppressed nonhydrodynamic sectors through the mode ratio $R=y_-/y_+$. Thus, the vanishing of the stokes constant does not imply the disappearance of the nonhydrodynamic transient mode or its nonlinear contributions to the transseries.

\section{Summary}
\label{sec:summary}

In this work, we investigated the projective structure of the normalized shear stress in Bjorken expanding relativistic viscous hydrodynamics across both the causal and acausal domains. 
Through a logarithmic-derivative transformation, the nonlinear Riccati equation was mapped to a second order linear differential equation. 
This reformulation allowed the hydrodynamic branch and the nonhydrodynamic transient contributions to be constructed from the two fundamental solutions of the auxiliary linear problem.

In contrast to the decoupled spin density sector, one major difference is that the linearized second order differential equation governing the normalized shear stress does not have a direct physical interpretation.
Although the auxiliary linear equation contains only two independent solutions, the nonlinear transformation back to the shear stress generates an infinite hierarchy of exponentially suppressed sectors. 
The leading nonperturbative sector is associated with the fundamental transient mode, while the higher sectors arise as nonlinear descendants generated by successive powers of the same mode ratio. 
These higher sectors therefore do not represent additional independent dynamical modes.

At late times, the projective representation also makes the attractor behavior transparent. The zeroth sector of the transseries corresponds to the hydrodynamic asymptotic branch, while the higher sectors describe nonlinear contributions generated by the transient mode. As the transient mode ratio becomes exponentially suppressed, solutions with finite transient amplitudes asymptotically lose their initial-condition dependence and approach the hydrodynamic branch.
The analysis also shows that convergence of the late time expansion and linear causality are distinct properties. 
The special convergent example lies outside the standard causal and stable regime, but generic acausal parameter choices remain factorially divergent. 
Convergence is therefore controlled by an algebraic termination condition rather than by the causal bound itself.
For generic parameter values, the hydrodynamic gradient expansion exhibits factorial growth and is therefore asymptotic. 
The Borel analysis of the auxiliary linear equation shows that this factorial growth originates from a Borel singularity, while the corresponding resurgence relation connects the large order coefficients of the algebraic auxiliary branch to the low-order coefficients of the exponentially suppressed transient branch.
At the discrete values $a=N^2$, with $N$ being a positive integer, the asymptotic expansions of the auxiliary fundamental solutions terminate. The corresponding hydrodynamic branch becomes a rational function with a convergent late time gradient expansion. At these same parameter values, the exact Stokes constant vanishes, indicating the disappearance of Stokes-induced coupling between the hydrodynamic and transient branches, although the transient mode itself remains present.

Whether the particularly transparent branch coupling structure found here is mainly a consequence of the simplified solvable model, or reflects a more general mechanism underlying hydrodynamic large order behavior, deserves further investigation.

\section*{Acknowledgements} 
This research is supported by the National Natural Science Foundation of China with Project Nos.~12635009, 12675175.

\bibliography{refs}
\end{document}